\documentclass[
aps,prd,preprint,letterpaper,
superscriptaddress,nofootinbib,
amsmath,amssymb,longbibliography
]{revtex4-2}
\usepackage{array}[=2016-10-06]
\usepackage{multirow}
\usepackage{diagbox}
\usepackage{graphicx}
\usepackage{bm}
\usepackage{slashed}
\usepackage{booktabs}
\usepackage{xcolor}
\usepackage[hidelinks]{hyperref}
\allowdisplaybreaks[4]

\newcommand{\as}{\alpha_s}

\DeclareMathOperator{\Rea}{Re}

\newcommand{\Tc}{T_{4c}}
\newcommand{\Tb}{T_{4b}}
\newcommand{\TQ}{T_{4Q}}

\newcommand{\Cthree}{C_{\bar{\bm{3}}\otimes\bm{3}}}
\newcommand{\Csix}{C_{\bm{6}\otimes\bar{\bm{6}}}}
\newcommand{\Othree}{\mathcal O_{\bar{\bm{3}}\otimes\bm{3}}}
\newcommand{\Osix}{\mathcal O_{\bm{6}\otimes\bar{\bm{6}}}}

\newcolumntype{C}[1]{>{\centering\arraybackslash}p{#1}}
\newcolumntype{R}[1]{>{\raggedleft\arraybackslash}p{#1}}
\newcolumntype{L}[1]{>{\raggedright\arraybackslash}p{#1}}

\begin{document}
\raggedbottom

\title{\mbox{}\\[10pt]
Next-to-leading-order QCD corrections to radiative decays
of fully heavy tetraquarks into quarkonia}
\author{Yu-Dong Zhang}
\email{ydzhang@mails.ccnu.edu.cn}

\author{Zai-Fu Jiang}
\email{200707008@jcut.edu.cn}

\affiliation{Jingchu University of Technology, Jingmen, Hubei 448000, China}
\begin{abstract}
Within nonrelativistic QCD factorization, we calculate the
next-to-leading-order (NLO) QCD corrections to the decays of
compact $T_{4Q}=QQ\bar Q\bar Q$ states ($Q=c,b$) into an
$S$-wave quarkonium and a photon. Conservation of charge conjugation allows $T_{4Q}^{0^{++}}$
and $T_{4Q}^{2^{++}}$ to decay into $V_Q+\gamma$, with
$V_c=J/\psi$ and $V_b=\Upsilon$, while $T_{4Q}^{1^{+-}}$
can decay into $\eta_Q+\gamma$. At leading order in the heavy quark velocity expansion, we derive
the short-distance coefficients for these decay widths in analytic
form through $\mathcal{O}(\alpha\alpha_s^5)$, including the
interference contribution between the two color configurations
of $T_{4Q}^{0^{++}}$. All ultraviolet divergences at NLO are exactly canceled by renormalization. The NLO QCD corrections enhance the decay widths for $T_{4Q}^{2^{++}}\to V_Q+\gamma$ and $T_{4Q}^{1^{+-}}\to\eta_Q+\gamma$ by factors of approximately $2.94$ ($2.33$) and $2.22$ ($1.83$), respectively. The dependence on the renormalization scale $\mu_R$ is
reduced at NLO in most channels, although it remains sizable. Based on the NLO branching fractions, we estimate yields of order
$10^6$--$10^7$ for $T_{4c}\to J/\psi+\gamma$ and $10^4$ for
$T_{4c}\to\eta_c+\gamma$ in $pp$ collisions at
$\sqrt{s}=13~\mathrm{TeV}$ with an integrated luminosity of
$100~\mathrm{fb}^{-1}$. The $J/\psi+\gamma$ channels can be investigated through $J/\psi\to\ell^+\ell^-$ in the $\ell^+\ell^-\gamma$ invariant-mass spectrum, complementing di-charmonium searches.
\end{abstract}
\maketitle
\newpage

\section{Introduction}\label{sec:intro}
In 2020, the LHCb Collaboration observed the narrow structure
$X(6900)$ and a broad near-threshold enhancement in the
di-$J/\psi$ invariant-mass spectrum~\cite{LHCb:2020bwg}.
CMS and ATLAS confirmed the $X(6900)$ signal and reported
additional structures in di-charmonium final
states~\cite{CMS:2023owd,ATLAS:2023bft}.
Using a larger data sample, CMS observed the $X(6600)$,
$X(6900)$, and $X(7100)$ structures and their mutual
interference with significances above
$5\sigma$~\cite{CMS:2026tiu}.
Assuming common quantum numbers for these states, CMS found
that their decay angular distributions favor a
$J^{PC}=2^{++}$ assignment~\cite{CMS:2025fpt}.
These findings motivate further studies of the production
and decay properties of fully charmed tetraquarks to clarify
their internal structure.

Radiative decays into an $S$-wave heavy quarkonium state and a
photon provide channels complementary to di-quarkonium decays.
In the $J/\psi\gamma$ and $\Upsilon\gamma$ channels, the
quarkonia can be reconstructed through their dilepton decays.
In particular, the $J/\psi\gamma$ channel can remain
kinematically open even for tetraquark masses below the
$J/\psi J/\psi$ threshold~\cite{Yang:2026bjy}.
Charge-conjugation conservation allows tetraquarks with
$C=+1$ and $C=-1$ to decay radiatively into vector and
pseudoscalar quarkonia, respectively.

For compact fully heavy tetraquarks, the heavy-quark mass $m_Q$
sets the hard scale for annihilation, well separated from the
nonrelativistic bound-state scales.
This scale separation motivates extending the nonrelativistic
QCD (NRQCD) factorization framework, originally developed for
quarkonium~\cite{Bodwin:1994jh}, to these systems.
In this framework, short-distance coefficients (SDCs) are
calculated perturbatively, while long-distance matrix elements
(LDMEs) encode the nonperturbative bound-state dynamics.
Applications include hadroproduction~\cite{Feng:2023agq},
parton fragmentation~\cite{Feng:2020riv,Bai:2024ezn},
photoproduction~\cite{Feng:2023ghc}, and tetraquark production
in association with a photon in $e^+e^-$
annihilation~\cite{Feng:2020qee,Liang:2025wbt}.
Electromagnetic and hadronic decays have also been
studied~\cite{Sang:2023ncm,Zhang:2023ffe}.
Recently, next-to-leading-order (NLO) QCD corrections have
been calculated for the hadroproduction~\cite{Wang:2025hex},
two-photon decays~\cite{Liu:2025mxv}, and
light-hadron decays~\cite{Wang:2026asl}
of fully charmed tetraquarks.

The radiative decays $\Tc\to J/\psi(\eta_c)+\gamma$ and
$\Tb\to\Upsilon(\eta_b)+\gamma$ were studied at leading order
(LO) by Yang \textit{et al.}~\cite{Yang:2026bjy}.
Extending these calculations to NLO provides a consistency
check of NRQCD factorization for tetraquark-to-quarkonium
transitions.
This involves checking whether all divergences in the NLO
amplitudes cancel after renormalization.

Previous studies of radiative quarkonium decays have found
sizable NLO QCD corrections whose magnitude and sign depend
on the decay channel.
For $\Upsilon\to\eta_c+\gamma$, the width increases to
approximately $1.5$ times its LO value, whereas that for
$\Upsilon\to\chi_{c2}+\gamma$ decreases by about
$40\%$~\cite{Zhang:2021ted}.
The NLO branching fraction for $\eta_b\to J/\psi+\gamma$
is approximately $3.65$ times the LO
result~\cite{Yang:2026kvh}.
The width for $\chi_{b0}\to J/\psi+\gamma$ is enhanced by
a factor of about $2.8$, and that for
$\chi_{b1}\to J/\psi+\gamma$ by about $1.8$.
In contrast, the $\chi_{b2}\to J/\psi+\gamma$ width is reduced
by approximately $36\%$~\cite{Zhang:2026chibgamma}.
For the tetraquark decays considered here, the leading QCD
contribution to the widths is of order $\alpha\alpha_s^4$,
making the LO predictions sensitive to the renormalization scale.
An NLO calculation is therefore needed to quantify the QCD
corrections and assess whether the scale dependence is reduced.

Electromagnetic contributions, including their interference
with QCD amplitudes, are numerically important in
$\Upsilon\to\eta_c+\gamma$ and
$\chi_{b0,2}\to J/\psi+\gamma$~\cite{Zhang:2021ted,Zhang:2026chibgamma}.
For the tetraquark decays considered here, the leading QCD
amplitudes arise at tree level and are of order
$\alpha^{1/2}\alpha_s^2$.
Additional electromagnetic amplitudes start at order
$\alpha^{3/2}\alpha_s$.
By coupling power counting, their squared contributions and
interference with the QCD amplitudes are suppressed relative
to the leading QCD widths by $(\alpha/\alpha_s)^2$ and
$\alpha/\alpha_s$, respectively.
We work at LO in $\alpha$ and neglect these higher-order
electromagnetic contributions.

In this work, we calculate the NLO QCD corrections to these
radiative decays of compact $\TQ=QQ\bar Q\bar Q$ states
with $Q=c,b$.
We consider the $J^{PC}=0^{++},1^{+-},2^{++}$ states and
work at LO in the heavy-quark relative velocities for both
the initial tetraquark and the final quarkonium.
For $\TQ^{0^{++}}$, we include the
$\bar{\mathbf{3}}\otimes\mathbf{3}$ and
$\mathbf{6}\otimes\bar{\mathbf{6}}$ color components
and their interference.
All divergences cancel in the NLO SDCs after renormalization
and matching, consistent with NRQCD factorization at this order.
We derive analytic expressions for the LO and NLO SDCs,
keeping the renormalization-scale dependence explicit,
and evaluate the decay widths using representative sets of LDMEs.
To assess the prospects for experimental observation, we also
estimate the yields of $\Tc\to J/\psi+\gamma$ and
$\Tc\to\eta_c+\gamma$ at the LHC and discuss their implications
for tetraquark searches.

The paper is organized as follows.
Section~\ref{sec:theory} presents the NRQCD factorization formulas
and the matching procedure.
The perturbative calculation and the SDCs through NLO are given
in Sec.~\ref{sec:SDCs}.
We discuss the phenomenological results in
Sec.~\ref{sec:phenomenology} and summarize our findings in
Sec.~\ref{sec:summary}.

\section{Theoretical framework}\label{sec:theory}
\subsection{NRQCD factorization for \texorpdfstring{$T_{4Q}$}{T4Q} decay}
\label{sec:factorization}

We consider compact $S$-wave tetraquarks in the
diquark--antidiquark basis.
Fermi statistics requires the spin-singlet $QQ$ diquark
to be in the color sextet and the spin-triplet diquark
to be in the color antitriplet.
The allowed quantum numbers are $J^{PC}=0^{++},1^{+-},2^{++}$.
Charge-conjugation conservation allows the radiative decays
\begin{equation}
	\TQ^{0^{++},\,2^{++}}\to V_Q+\gamma,
	\qquad
	\TQ^{1^{+-}}\to\eta_Q+\gamma,
	\label{eq:channels}
\end{equation}
where $V_c=J/\psi$, $V_b=\Upsilon$, $\eta_c$ and $\eta_b$
are the corresponding pseudoscalar quarkonia.
Both the $\bar{\bm{3}}\otimes\bm{3}$ and
$\bm{6}\otimes\bar{\bm{6}}$ color configurations are allowed
for $\TQ^{0^{++}}$, whereas only the
$\bar{\bm{3}}\otimes\bm{3}$ configuration contributes to
$\TQ^{1^{+-}}$ and $\TQ^{2^{++}}$ at LO in the velocity
expansion~\cite{Sang:2023ncm,Yang:2026bjy,Huang:2021vtb}.

After integration over the two-body phase space, the unpolarized
decay width reads
\begin{equation}
	\Gamma[\TQ^{(J)}\to H+\gamma]
	=
	\frac{|\boldsymbol k|}
	{8\pi m_{T_{4Q}}^2(2J+1)}
	\sum_{\mathrm{pol.}}
	\left|\mathcal M[\TQ^{(J)}\to H+\gamma]\right|^2,
	\label{eq:physical-width}
\end{equation}
where $\mathcal M$ is the physical decay amplitude,
with $H=V_Q$ for $J=0,2$ and $H=\eta_Q$ for $J=1$.
The sum runs over the initial- and final-state polarizations,
and the factor $1/(2J+1)$ accounts for the initial-state
spin average.
The magnitude of the photon three-momentum in the tetraquark
rest frame is
\begin{equation}
	|\boldsymbol k|
	=\frac{m_{T_{4Q}}^2-m_H^2}{2m_{T_{4Q}}},
	\label{eq:physical-photon-momentum}
\end{equation}
where $m_{T_{4Q}}$ and $m_H$ are the tetraquark and quarkonium
masses, respectively.
At LO in the velocity expansion, we set $m_{T_{4Q}}=4m_Q$
and $m_H=2m_Q$ in both the short-distance calculation
and the phase-space factor, where $m_Q$ is the heavy-quark
pole mass.
We retain terms through $\mathcal{O}(\alpha\as^5)$
in the squared amplitude.
The phase-space factor and the initial-state spin average
are kept explicit in the factorization formulas below
and are not included in the SDCs.

Adapting the factorization formulas of
Refs.~\cite{Sang:2023ncm,Yang:2026bjy}, with the phase-space
and LDME conventions specified here, we write
\begin{subequations}\label{eq:factorization}
\begin{align}
 \Gamma[\TQ^{0^{++}}\to V_Q\gamma]
 =&\frac{|\boldsymbol k|}
 {8\pi m_{T_{4Q}}^2}\big(
 c_1\langle\Osix^{(0)}\rangle
 +c_2\langle\Othree^{(0)}\rangle
 +2c_{\mathrm{mix}}\langle\mathcal O_{\mathrm{mix}}^{(0)}\rangle
 \big){}\langle \mathcal O_{V_Q}\rangle,\label{eq:width0}\\
 \Gamma[\TQ^{2^{++}}\to V_Q\gamma]
 =&\frac{|\boldsymbol k|}
 {40\pi m_{T_{4Q}}^2}{}c_3\langle\Othree^{(2)}\rangle\langle \mathcal O_{V_Q}\rangle,
 \label{eq:width2}\\
 \Gamma[\TQ^{1^{+-}}\to\eta_Q\gamma]
 =&\frac{|\boldsymbol k|}
 {24\pi m_{T_{4Q}}^2}{}c_4\langle\Othree^{(1)}\rangle\langle \mathcal O_{\eta_Q}\rangle.
 \label{eq:width1}
\end{align}
\end{subequations}
Here $c_1,c_2,c_3,c_4$ and $c_{\mathrm{mix}}$ are the SDCs. The normalization factors
associated with the external states are included in these coefficients.
The tetraquark LDMEs use the spin-summed convention defined below. The initial-state spin averages
remain explicit in Eq.~\eqref{eq:factorization}.
The tetraquark and quarkonium LDMEs have dimensions $9$ and $3$, respectively.
Relativistic corrections to Eq.~\eqref{eq:factorization} are not included.

The tetraquark LDMEs are defined by
\begin{subequations}\label{eq:T-LDMEs}
	\begin{align}
		\langle\Osix^{(0)}\rangle
		&=
		\left|\langle0|\Osix^{(0)}|\TQ^{0^{++}}\rangle\right|^2,
		\\
		\langle\Othree^{(J)}\rangle
		&=
		\sum_{\lambda=-J}^{J}
		\left|\langle0|\Othree^{(J)}(\lambda)
		|\TQ^{(J)}(\lambda)\rangle\right|^2,
		\\
		\langle\mathcal O_{\mathrm{mix}}^{(0)}\rangle
		&=
		\Rea\!\left[
		\langle0|\Othree^{(0)}|\TQ^{0^{++}}\rangle
		\langle\TQ^{0^{++}}|\Osix^{(0)\dagger}|0\rangle
		\right],
		\label{eq:mix-LDME}
	\end{align}
\end{subequations}
where $J=0,1,2$, and the tetraquark states are
nonrelativistically normalized.
The operator $\Othree^{(J)}(\lambda)$ is projected onto
the polarization $\lambda$ of the tetraquark state.
For $J=0$, the polarization label is omitted.
The diagonal LDMEs include a sum over the $2J+1$ spin
polarizations and are therefore $2J+1$ times the corresponding
fixed-spin decay LDMEs in Table~I of
Ref.~\cite{Sang:2023ncm}.
Within the vacuum-saturation approximation, these definitions
reproduce the spin-summed production LDME normalization
of Ref.~\cite{Feng:2023agq}.
The mixed LDME need not be positive.
The relative phase convention for the two color operators
must be used consistently in this LDME and $c_{\mathrm{mix}}$.

Using the diquark operators of
Refs.~\cite{Sang:2023ncm,Huang:2021vtb}, with the polarization
contractions made explicit, we define
\begin{subequations}\label{eq:operators}
\begin{align}
 \Osix^{(0)}
 &=[\psi_a^T(i\sigma^2)\psi_b]
   [\chi_c^\dagger(i\sigma^2)\chi_d^*]\Csix^{ab;cd},\\
 \Othree^{(0)}
 &=-\frac1{\sqrt3}[\psi_a^T(i\sigma^2)\sigma^i\psi_b]
   [\chi_c^\dagger\sigma^i(i\sigma^2)\chi_d^*]\Cthree^{ab;cd},\\
 \Othree^{(1)}(\lambda)
 &=-\frac{i}{\sqrt2}\epsilon_T^{i*}(\lambda)\varepsilon^{ijk}
   [\psi_a^T(i\sigma^2)\sigma^j\psi_b]
   [\chi_c^\dagger\sigma^k(i\sigma^2)\chi_d^*]\Cthree^{ab;cd},\\
 \Othree^{(2)}(\lambda)
 &=\frac12\epsilon_T^{ij*}(\lambda)\Gamma^{ij;kl}
   [\psi_a^T(i\sigma^2)\sigma^k\psi_b]
   [\chi_c^\dagger\sigma^l(i\sigma^2)\chi_d^*]\Cthree^{ab;cd}.
\end{align}
\end{subequations}
Here $\psi$ and $\chi^\dagger$ annihilate the heavy quark and antiquark,
respectively, and $\sigma^i$ are Pauli matrices. The indices
$i,j,k,l$ are spatial indices and $a,b,c,d$ are color indices. The
polarization vector and tensor of the tetraquark are denoted by
$\epsilon_T^i(\lambda)$ and $\epsilon_T^{ij}(\lambda)$, respectively. The symmetric traceless tensor and the color tensors are
\begin{align}
 \Gamma^{ij;kl}&=\delta^{ik}\delta^{jl}+\delta^{il}\delta^{jk}-\frac23\delta^{ij}\delta^{kl},\\
 \Cthree^{ab;cd}&=\frac{\delta^{ac}\delta^{bd}-\delta^{ad}\delta^{bc}}{2\sqrt3},\\
 \Csix^{ab;cd}&=\frac{\delta^{ac}\delta^{bd}+\delta^{ad}\delta^{bc}}{2\sqrt6}.
 \label{eq:colors}
\end{align}
The two color tensors have unit norm and are orthogonal.

For the final quarkonium, we adopt the color-singlet
LDME convention of Ref.~\cite{Yang:2026bjy}. At LO in the velocity
expansion, the LDMEs are given by~\cite{Bodwin:1994jh}
\begin{equation}
	\langle \mathcal O_{V_Q}\rangle=\frac{3N_c}{2\pi}|R_{V_Q}(0)|^2,
	\qquad
	\langle \mathcal O_{\eta_Q}\rangle=\frac{N_c}{2\pi}|R_{\eta_Q}(0)|^2,
	\label{eq:H-LDMEs}
\end{equation}
where $N_c=3$.

In the convention of Eq.~\eqref{eq:T-LDMEs}, the LDMEs
are related to the potential-model wave functions
by~\cite{Feng:2023agq,Sang:2023ncm}
\begin{align}
	\langle\Othree^{(J)}\rangle
	&\approx16(2J+1)|\psi_3^{(J)}(0)|^2,\nonumber\\
	\langle\Osix^{(0)}\rangle
	&\approx16|\psi_6^{(0)}(0)|^2,\nonumber\\
	\langle\mathcal O_{\mathrm{mix}}^{(0)}\rangle
	&\approx16\Rea\!\left[
	\psi_3^{(0)}(0)\psi_6^{(0)*}(0)
	\right].
	\label{eq:T-wavefunctions}
\end{align}
Here $\psi_3^{(J)}$ and $\psi_6^{(0)}$ denote the
$\bar{\mathbf{3}}\otimes\mathbf{3}$ and
$\mathbf{6}\otimes\bar{\mathbf{6}}$ components of the
tetraquark wave function, respectively, in the operator
basis of Eq.~\eqref{eq:operators}.
The mixed LDME retains the relative phase of the two components.
For a physical $\TQ^{0^{++}}$ state, these components already
include the color-mixing, so no additional mixing angle
enters Eq.~\eqref{eq:width0}.

\subsection{Procedure to determine the SDCs}\label{sec:matching}
The SDCs are insensitive to long-distance bound-state dynamics
and are determined by perturbative matching between QCD and NRQCD. We replace $\TQ$ by a free, on-shell $QQ\bar Q\bar Q$ state $\widetilde T_{4Q}$ in the relevant spin and color configuration, and the quarkonium by a free, on-shell color-singlet $Q\bar Q$ state $\widetilde H$ with the corresponding spin. Both sides of Eq.~\eqref{eq:factorization}, including the perturbative NRQCD matrix elements, are expanded through NLO in $\as$ to extract the SDCs.

Let $P$, $p$, and $k$ denote the tetraquark, quarkonium, and photon
momenta. At leading order in the velocity expansion,
\begin{align}
 P=p+k,\qquad P^2=16m_Q^2,\qquad p^2=4m_Q^2,\qquad k^2=0,\qquad p\cdot k=6m_Q^2.
 \label{eq:kinematics}
\end{align}
The phase-space factor becomes
\begin{equation}
 \frac{|\boldsymbol k|}{8\pi m_{T_{4Q}}^2}
 =\frac{3}{256\pi m_Q}.
 \label{eq:phase-space-kinematics}
\end{equation}

We generate the partonic amplitude for
$QQ\bar Q\bar Q\to Q\bar Q+\gamma$ with open Dirac
and color indices.
Following Refs.~\cite{Sang:2023ncm,Huang:2021vtb},
we implement the initial-state spin and color projections as
\begin{subequations}\label{eq:initial-projections}
	\begin{align}
		u_i^a u_j^b\bar v_k^c\bar v_l^d
		&\longrightarrow
		(C\Pi_0)_{ij}(\Pi_0 C)_{lk}\Csix^{ab;cd},
		\label{eq:project66}\\
		u_i^a u_j^b\bar v_k^c\bar v_l^d
		&\longrightarrow
		(C\Pi_1^\mu)_{ij}(\Pi_1^\nu C)_{lk}
		\Cthree^{ab;cd}\epsilon^{(J)}{\mu\nu}.
		\label{eq:project33}
	\end{align}
\end{subequations}
Here $i,j,k,l$ are Dirac indices and
$C$ is the charge-conjugation matrix.
The momentum argument $P/4$ of each initial-state spinor
is suppressed.
For the spinor ordering in Eq.~\eqref{eq:initial-projections},
the spin projectors are
\begin{equation}
	\Pi_0=\frac{(\slashed P/4-m_Q)\gamma_5}{\sqrt2},
	\qquad
	\Pi_1^\mu=\frac{(\slashed P/4-m_Q)\gamma^\mu}{\sqrt2}.
	\label{eq:diquark-projectors}
\end{equation}
The first replacement selects the
$\mathbf6\otimes\bar{\mathbf6}$ component of $\TQ^{0^{++}}$.
In the second, $\epsilon^{(J)}_{\mu\nu}$ couples the spin-triplet
diquark and antidiquark to total spin $J=0,1,2$.
It denotes the lower-index form of
$\epsilon^{(J)\mu\nu}(\lambda)$ introduced below,
with the polarization label $\lambda$ suppressed.

For the final color-singlet quarkonium, we use
\begin{equation}
	v(p/2)\bar u(p/2)\longrightarrow
	\frac{(\slashed p/2-m_Q)\Gamma_H}{\sqrt{2 N_c}},
	\label{eq:quarkonium-projector}
\end{equation}
where $\Gamma_{\eta_Q}=\gamma_5$ and
$\Gamma_{V_Q}=\slashed\epsilon_H^*$,
with $p\cdot\epsilon_H=0$.
The color identity matrix in
Eq.~\eqref{eq:quarkonium-projector} is implicit.
The final-state color-singlet factor is retained separately
from the tetraquark color tensor.

For quark-rearrangement diagrams, we apply the spin and color
projectors to the full amplitude with open external indices,
including the fermion lines connecting the initial constituents
to the final quarkonium.
The resulting spinor contractions are expressed as Dirac traces
using charge-conjugation identities~\cite{Yang:2026bjy}.
For a Dirac matrix chain $X$, we define
$X^C\equiv C^{-1}X^T C$, where $T$ denotes transposition
in Dirac space. The relevant identities are
\[
C^{-1}(\gamma^\mu)^T C=-\gamma^\mu,
\qquad
(X_1X_2)^C=X_2^C X_1^C.
\]
These identities determine the reversal of matrix order.
The relative signs associated with identical-fermion
permutations are retained throughout the calculation.

In the one-loop matching calculation, we perform the spin contractions
in $D=4-2\epsilon$ dimensions. For the
$\bar{\mathbf{3}}\otimes\mathbf{3}$ configuration, the two indices
of the polarization tensor $\epsilon^{(J)\mu\nu}(\lambda)$ refer to
the spins of the diquark and antidiquark, respectively. Their tensor
product decomposes into a trace ($J=0$), an antisymmetric tensor
($J=1$), and a symmetric traceless tensor ($J=2$), as in the
$D$-dimensional treatment of ${}^3P_J$ quarkonium in
Ref.~\cite{Petrelli:1997ge}. The scalar
$\mathbf{6}\otimes\bar{\mathbf{6}}$ configuration, in contrast,
is formed from two spin singlets and requires no rank-two
polarization tensor.

Defining the tensor transverse to the tetraquark momentum $P$ by
\[
\eta^{\mu\nu}=-g^{\mu\nu}+\frac{P^\mu P^\nu}{P^2},
\]
we obtain the polarization sums
\[
\begin{aligned}
	\epsilon^{(0)\mu\nu}\epsilon^{(0)*\rho\sigma}
	&=\frac{1}{D-1}\eta^{\mu\nu}\eta^{\rho\sigma},
	\\[4pt]
	\sum_{\lambda}
	\epsilon^{(1)\mu\nu}(\lambda)\epsilon^{(1)*\rho\sigma}(\lambda)
	&=\frac{1}{2}
	\left(\eta^{\mu\rho}\eta^{\nu\sigma}
	-\eta^{\mu\sigma}\eta^{\nu\rho}\right),
	\\[4pt]
	\sum_{\lambda}
	\epsilon^{(2)\mu\nu}(\lambda)\epsilon^{(2)*\rho\sigma}(\lambda)
	&=\frac{1}{2}
	\left(\eta^{\mu\rho}\eta^{\nu\sigma}
	+\eta^{\mu\sigma}\eta^{\nu\rho}\right)
	-\frac{1}{D-1}\eta^{\mu\nu}\eta^{\rho\sigma}.
\end{aligned}
\]
Contracting each expression with $g_{\mu\rho}g_{\nu\sigma}$ gives
the corresponding polarization multiplicities,
\begin{equation}
	N_0(D)=1,\qquad
	N_1(D)=\frac{(D-1)(D-2)}{2},\qquad
	N_2(D)=\frac{(D-2)(D+1)}{2}.
	\label{eq:spin-multiplicities}
\end{equation}
In perturbative matching, the spin sum in the tetraquark LDME runs
over $N_J(D)$ polarizations, while the average over the initial
tetraquark spin carries a factor of $1/N_J(D)$. In the
four-dimensional limit, $N_J(D)$ reduces to $2J+1$.

For a final-state vector quarkonium of momentum $p$, we use
\[
\sum_{\lambda_H}
\epsilon_H^{\alpha *}(\lambda_H)\epsilon_H^\beta(\lambda_H)
=-g^{\alpha\beta}+\frac{p^\alpha p^\beta}{p^2}.
\]
For the outgoing photon, the electromagnetic Ward identity allows
us to replace the physical polarization sum by $-g^{\delta\delta'}$
in the squared amplitude once all diagrams have been combined.
We retain the full $D$ dependence of the spin projectors,
polarization sums, spin averages, and matching normalization
factors until renormalization and NRQCD matching are complete.

For the scalar and tensor configurations, the tree-level
perturbative LDME for a normalized polarization is
$4^2(2m_Q)^4$ when the heavy-quark states are relativistically
normalized~\cite{Sang:2023ncm}. The axial-vector matrix element
in Eq.~(8) of Ref.~\cite{Huang:2021vtb} gives the same fixed-polarization
factor after conversion to this external-state normalization.
In the spin-summed convention of Eq.~\eqref{eq:T-LDMEs}, the
diagonal perturbative normalization is consequently $N_J 4^2(2m_Q)^4$.
For the scalar sextet configuration and the scalar interference
term, $N_0=1$. The factor $N_J$ is included when dividing
out the perturbative LDME in the matching.
\section{Perturbative calculation and the SDCs}\label{sec:SDCs}
\begin{figure}[h]
	\centering
	\includegraphics[width=0.92\textwidth]
	{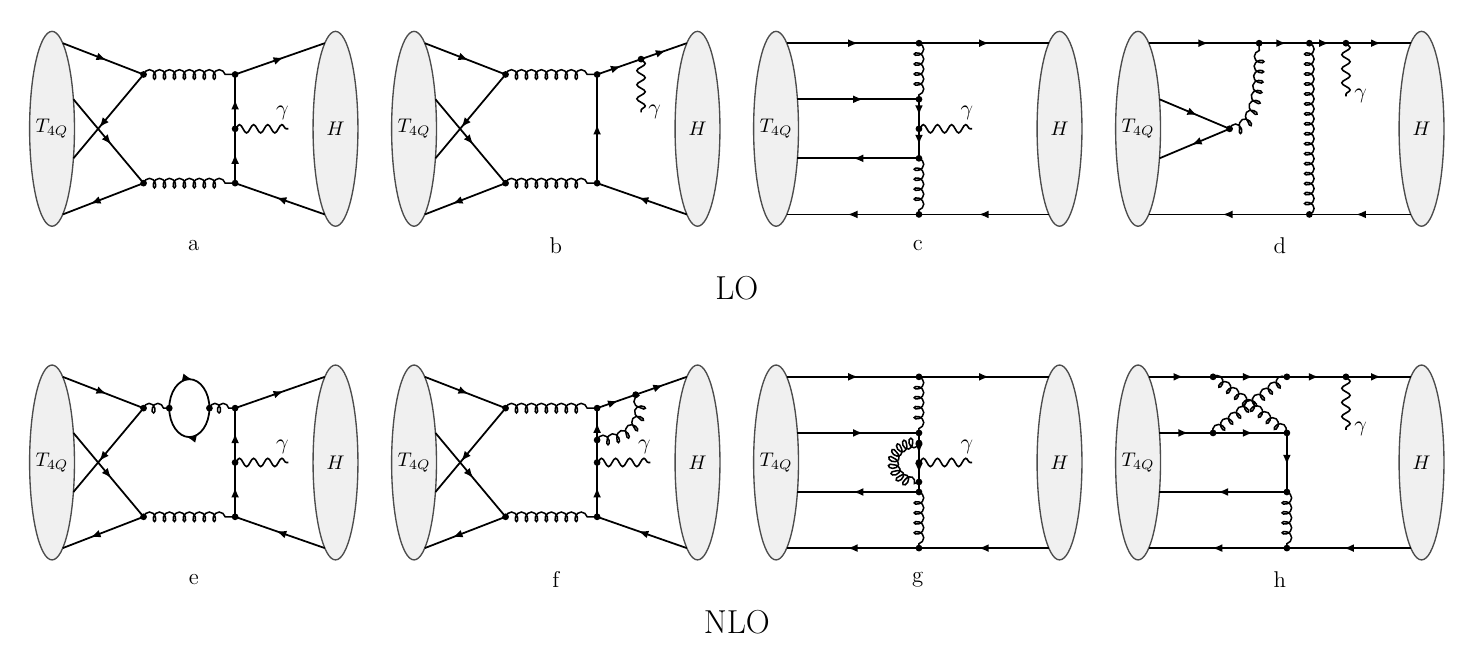}
	\caption{Representative tree-level and one-loop Feynman diagrams for \(T_{4Q}\to V_Q+\gamma\) and \(T_{4Q}\to\eta_Q+\gamma\), with \(Q=c,b\), drawn with JaxoDraw~\cite{Binosi:2003yf}.}
	\label{fig-feynman-diagrams-t4q}
\end{figure}
For the QCD contribution, we perform the perturbative matching using
the quark-level process $QQ\bar Q\bar Q\to Q\bar Q+\gamma$ with
$Q=c,b$. The initial four-quark states are projected onto the allowed
spin and color configurations of $T_{4Q}$, and the final quark pair
onto a color-singlet ${}^3S_1$ or ${}^1S_0$ state.
The tree-level amplitudes and their one-loop QCD corrections are
of order $\alpha^{1/2}\alpha_s^2$ and $\alpha^{1/2}\alpha_s^3$,
respectively. The diagrams and amplitudes are generated with
\textsc{FeynArts}~\cite{Hahn:2000kx}, yielding 288 tree and
11340 one-loop diagrams. Representative diagrams are shown
in Fig.~\ref{fig-feynman-diagrams-t4q}.

At LO in the velocity expansion, we set the heavy-quark relative
momenta to zero before loop integration. Within the method of regions, this procedure isolates the hard contribution to the SDCs~\cite{Beneke:1997zp}. The Dirac and color algebra is performed with \textsc{FeynCalc}~\cite{Mertig:1990an,Shtabovenko:2016sxi}, together with
\textsc{FormLink}~\cite{Feng:2012tk} and
\textsc{FORM}~\cite{Kuipers:2012rf}.
The interference contributions are expressed as linear combinations
of scalar one-loop integrals upon summation over polarizations.
After partial-fraction decomposition with
\textsc{Apart}~\cite{Feng:2012iq}, these integrals are reduced
to master integrals using integration-by-parts
identities~\cite{Tkachov:1981wb,Chetyrkin:1981qh}, implemented in
\textsc{FIRE}~\cite{Smirnov:2014hma}.
The master integrals are evaluated analytically with
\textsc{Package-X}~\cite{Patel:2015tea}, and the results are
cross-checked numerically with \textsc{AMFlow}~\cite{Liu:2022chg}.

We use dimensional regularization in $D=4-2\epsilon$ dimensions
to regulate both ultraviolet (UV) and infrared (IR) divergences.
For $Q=c,b$, we renormalize the heavy-quark mass and field in the
on-shell scheme and the strong coupling $\alpha_s$ in the $\overline{\mathrm{MS}}$ scheme.
Writing $Z_i=1+\delta Z_i$, we obtain the one-loop
counterterms~\cite{Zhang:2005cha}:
\begin{align}
	\delta Z_m^{\mathrm{OS}}
	&=-3C_F\frac{\alpha_s}{4\pi}
	\left[
	\frac{1}{\epsilon_{\mathrm{UV}}}
	-\gamma_E+\ln\frac{4\pi\mu_R^2}{m_Q^2}
	+\frac{4}{3}
	\right],\nonumber\\
	\delta Z_2^{\mathrm{OS}}
	&=-C_F\frac{\alpha_s}{4\pi}
	\left[
	\frac{1}{\epsilon_{\mathrm{UV}}}
	+\frac{2}{\epsilon_{\mathrm{IR}}}
	-3\gamma_E+3\ln\frac{4\pi\mu_R^2}{m_Q^2}
	+4
	\right],\nonumber\\
	\delta Z_{\alpha_s}^{\overline{\mathrm{MS}}}
	&=-\beta_0\frac{\alpha_s}{4\pi}
	\left[
	\frac{1}{\epsilon_{\mathrm{UV}}}
	-\gamma_E+\ln(4\pi)
	\right].
	\label{eq:counterterms}
\end{align}
Here $\beta_0=(11C_A-4T_F n_f)/3$ is the one-loop coefficient of the QCD beta function, where $n_f$ is the number of active quark
flavors in the $\overline{\mathrm{MS}}$ coupling. For
$\mathrm{SU}(3)_c$, the color factors are $C_A=3$, $C_F=4/3$,
and $T_F=1/2$. After renormalization, matching the one-loop QCD amplitudes onto NRQCD yields SDCs that are free of both UV and IR poles. As checks of the calculation, we verify the electromagnetic Ward identity for the projected amplitudes and reproduce the tree-level results for $T_{4c}$ in Appendix~A of Ref.~\cite{Yang:2026bjy} after accounting for the different normalization conventions.

Through NLO in $\alpha_s$, the SDCs obtained by perturbative
matching take the form
\begin{equation}
	c_i(\mu_R)
	= \frac{\pi^5 e_Q^2 \alpha \alpha_s^4}{m_Q^{10}}
	\left[
	c_i^{(0)}
	+ \frac{\alpha_s}{\pi}
	\left(
	\beta_0 c_i^{(0)} \ln\frac{\mu_R^2}{4m_Q^2}
	+ c_i^{(1)}
	\right)
	\right]
	+ \mathcal{O}(\alpha_s^6),
	\label{eq:NLO-scale}
\end{equation}
where $m_Q$ is the heavy-quark
pole mass, and $e_Q$ is the electric charge of the heavy quark in units of the positron charge, with $e_c=2/3$ and $e_b=-1/3$.
The dimensionless coefficients $c_i^{(1)}$ are the finite NLO
terms at the reference scale $\mu_R=2m_Q$, with the pole masses
and the NRQCD factorization convention held fixed.
For $T_{4Q}^{0^{++}}\to V_Q+\gamma$, the indices $1$, $2$, and
$\mathrm{mix}$ label the contributions from the
$\bm{6}\otimes\bar{\bm{6}}$ and $\bar{\bm{3}}\otimes\bm{3}$
color configurations and their interference, respectively.
The indices $3$ and $4$ refer to the decays
$T_{4Q}^{2^{++}}\to V_Q+\gamma$ and
$T_{4Q}^{1^{+-}}\to\eta_Q+\gamma$, respectively.

At LO, the perturbative matching gives
\begin{subequations}\label{eq:LO-SDCs}
	\begin{align}
		c_1^{(0)}&=\frac{1155200}{729},\\
		c_2^{(0)}&=\frac{6149069056}{54675},\\
		c_{\mathrm{mix}}^{(0)}&=-\frac{11919232}{729}\sqrt{\frac23},\\
		c_3^{(0)}&=\frac{2885566976}{273375},\\
		c_4^{(0)}&=\frac{2471326208}{18225}.
	\end{align}
\end{subequations}
Relative to Eq.~(A1) in Appendix~A of
Ref.~\cite{Yang:2026bjy}, our LO coefficients contain an
overall factor of $1/4$. A previously missing factor of
$1/4$ in the diagonal SDCs was also restored in the revised
arXiv version of Ref.~\cite{Sang:2023ncm}. The tensor and axial-vector coefficients additionally include the inverse spin
multiplicities $1/5$ and $1/3$, respectively, in accordance
with the spin-summed tetraquark LDME convention used here.
The factor of 2 multiplying the interference term is kept
explicitly in Eq.~\eqref{eq:width0}.

At NLO, their analytic expressions are collected in
Appendix~\ref{app:finite-NLO-coefficients}.
Here, $n_l$ and $n_h$ count massless quark flavors and heavy
quark flavors of mass $m_Q$, respectively, while the factor $lbl$ labels the light-by-light
contributions in the normalization of the analytic expressions.
Its numerical values are $lbl=1$ for charm and $lbl=1/2$ for
bottom.
\begin{table}[htbp]
	\centering
	\caption{SDCs for the radiative decays of $T_{4Q}$.
		The tabulated values of $c_i^{(0)}$ and $c_i^{(1)}$
		are given in units of $10^4$.}
	\label{tab:SDCs-charm}
	\begingroup
	\small
	\renewcommand{\arraystretch}{0.8}
	\setlength{\tabcolsep}{16pt}
	\begin{tabular}{|c|c|c|c|}
		\hline
		Channel & Color & $c_i^{(0)}$ & $c_i^{(1)}$\\
		\hline
		
		& $\mathbf{6}\otimes\overline{\mathbf{6}}$
		& $0.16$
		& $-2.69-0.40n_l-0.03n_h+0.05\,\mathrm{lbl}$
		\\
		\cline{2-4}
		$T_{4Q}^{0^{++}}\to J/\psi(\Upsilon)+\gamma$
		& $\overline{\mathbf{3}}\otimes\mathbf{3}$
		& $11.25$
		& $327.86-26.18n_l-10.20n_h-0.35\,\mathrm{lbl}$
		\\
		\cline{2-4}
		
		& $\mathrm{mix}$
		& $-1.33$
		& $-8.14+3.22n_l+0.71n_h-0.20\,\mathrm{lbl}$
		\\
		\hline
		$T_{4Q}^{2^{++}}\to J/\psi(\Upsilon)+\gamma$
		& $\overline{\mathbf{3}}\otimes\mathbf{3}$
		& $1.06$
		& $34.99-1.94n_l-1.89n_h+0.03\,\mathrm{lbl}$
		\\
		\hline
		$T_{4Q}^{1^{+-}}\to \eta_Q+\gamma$
		& $\overline{\mathbf{3}}\otimes\mathbf{3}$
		& $13.56$
		& $289.92-18.30n_l-13.54n_h-0.11\,\mathrm{lbl}$
		\\
		\hline
	\end{tabular}
	\endgroup
\end{table}

For the phenomenological analysis, the numerical values of
the dimensionless coefficients $c_i^{(0)}$ and $c_i^{(1)}$
are summarized in Table~\ref{tab:SDCs-charm}. Since the common mass and coupling
factors have been extracted in Eq.~\eqref{eq:NLO-scale},
the tabulated coefficients are independent of $m_Q$ within
these approximations. Their flavor dependence is encoded
in $n_l$, $n_h$, and $ lbl $.
\section{Phenomenology and discussion}
\label{sec:phenomenology}
We take the heavy-quark pole masses to be
$m_c=1.5~\mathrm{GeV}$ and $m_b=4.9~\mathrm{GeV}$,
as in Ref.~\cite{Yang:2026bjy}, and set
$\alpha=1/137$~\cite{Sang:2023ncm}.
The strong coupling is evolved at one loop using
\texttt{RunDec}~\cite{Chetyrkin:2000yt}, with
\begin{equation}
	\alpha_s(2m_c)=0.236,\qquad
	\alpha_s(2m_b)=0.174.
	\label{eq:input-couplings}
\end{equation}
We choose $\mu_R=2m_Q$ as the central renormalization scale
and evaluate the scale variation at $\mu_R=m_Q$ and $4m_Q$,
keeping all other inputs fixed. 
For each decay channel, the color contributions are summed
at each scale before evaluating the variation. The kinematics are evaluated at
$m_{T_{4Q}}=4m_Q$ and $m_H=2m_Q$.

For the final-state quarkonia, we use the Buchm\"uller--Tye(B--T)
potential model wave functions in Tables~I and II of
Ref.~\cite{Eichten:1995ch},
\begin{align}
	\left|R_{1S}^{c\bar c}(0)\right|^2
	&=0.810~\mathrm{GeV}^3,
	\nonumber\\
	\left|R_{1S}^{b\bar b}(0)\right|^2
	&=6.477~\mathrm{GeV}^3.
	\label{eq:radial-wavefunction-inputs}
\end{align}
At leading order in the velocity expansion, heavy-quark spin
symmetry relates the vector and pseudoscalar wave functions
within each flavor sector~\cite{Bodwin:1994jh}.
With $N_c=3$ and the LDME conventions of Eq.~\eqref{eq:H-LDMEs},
we obtain
\begin{subequations}\label{eq:quarkonium-inputs}
	\begin{align}
		\langle\mathcal O_{J/\psi}\rangle
		&\approx\frac{3N_c}{2\pi}
		\left|R_{1S}^{c\bar c}(0)\right|^2
		=1.160~\mathrm{GeV}^3,
		\label{eq:LDME-Jpsi}\\
		\langle\mathcal O_{\eta_c}\rangle
		&\approx\frac{N_c}{2\pi}
		\left|R_{1S}^{c\bar c}(0)\right|^2
		=0.3867~\mathrm{GeV}^3,
		\label{eq:LDME-etac}\\
		\langle\mathcal O_{\Upsilon}\rangle
		&\approx\frac{3N_c}{2\pi}
		\left|R_{1S}^{b\bar b}(0)\right|^2
		=9.278~\mathrm{GeV}^3,
		\label{eq:LDME-Upsilon}\\
		\langle\mathcal O_{\eta_b}\rangle
		&\approx\frac{N_c}{2\pi}
		\left|R_{1S}^{b\bar b}(0)\right|^2
		=3.093~\mathrm{GeV}^3.
		\label{eq:LDME-etab}
	\end{align}
\end{subequations}

\begin{table*}[t]
	\centering
	\caption{Tetraquark LDME inputs in units of $\mathrm{GeV}^9$.
		The $J=1,2$ entries include the spin multiplicities $2J+1$.
		The mixed LDME excludes the factor of $2$ in the interference term.
		The entries for $T_{4b}^{0^{++}}$ in Sets~3 and~4 correspond to
		$0_{\mathrm I}^{++}$ and $0_{\mathrm{II}}^{++}$, respectively,
		with $\theta_b=66^\circ$.}
	\label{tab:tetraquark-LDMEs}
	\footnotesize
	\renewcommand{\arraystretch}{1.20}
	\setlength{\tabcolsep}{15pt}
	\begin{tabular}{|c|l|c|c|c|c|c|}
		\hline
		\multicolumn{2}{|c|}{$J^{PC}$}
		& \multicolumn{3}{c|}{$0^{++}$}
		& $1^{+-}$ & $2^{++}$ \\
		\hline
		\multicolumn{2}{|c|}{LDMEs}
		& $\langle O_{\mathbf6\otimes\bar{\mathbf6}}^{(0)}\rangle$
		& $\langle O_{\bar{\mathbf3}\otimes\mathbf3}^{(0)}\rangle$
		& $\langle O_{\mathrm{mix}}^{(0)}\rangle$
		& $\langle O_{\bar{\mathbf3}\otimes\mathbf3}^{(1)}\rangle$
		& $\langle O_{\bar{\mathbf3}\otimes\mathbf3}^{(2)}\rangle$ \\
		\hline
		\multirow{2}{*}{$T_{4c}$}
		& Set~1~\cite{Lu:2020cns,Sang:2023ncm}
		& $0.0128$ & $0.0347$ & $0.0211$ & $0.0780$ & $0.0720$ \\
		\cline{2-7}
		& Set~2~\cite{liu:2020eha,Sang:2023ncm}
		& $0.0139$ & $0.0187$ & $-0.0161$ & $0.0480$ & $0.0630$ \\
		\hline
		\multirow{3}{*}{$T_{4b}$}
		& Set~3~\cite{Yang:2026bjy}
		& $1.2318$ & $0.0528$ & $0.2549$ & $0.9590$ & $1.5990$ \\
		\cline{2-7}
		& Set~4~\cite{Yang:2026bjy}
		& $0.2442$ & $0.2662$ & $-0.2549$ & $0.9590$ & $1.5990$ \\
		\cline{2-7}
		& Set~5~\cite{Liang:2025wbt}
		& \textemdash & $0.3101$ & \textemdash & $0.9303$ & $1.5504$ \\
		\hline
	\end{tabular}
\end{table*}

Table~\ref{tab:tetraquark-LDMEs} lists the tetraquark LDMEs used
in this work. For $T_{4c}$, Sets~1 and~2 correspond to Models~I
and~II in Table~I of Ref.~\cite{Sang:2023ncm}, based on the
potential models of Refs.~\cite{Lu:2020cns,liu:2020eha},
respectively. In Eq.~\eqref{eq:T-LDMEs}, we sum over the
polarizations of the tetraquark. We therefore multiply the
$J=1$ and $J=2$ matrix elements of Ref.~\cite{Sang:2023ncm}
by $3$ and $5$, respectively. For $J=0$, we use the matrix
elements of each model without modification, including the sign
of the mixed LDME, since color mixing is already taken into
account. In Ref.~\cite{Yang:2026bjy}, the $T_{4b}$ matrix
elements in Table~III include the factor $2J+1$ in Eq.~(11),
whereas the $J=1,2$ matrix elements for $T_{4c}$ in Table~I
are taken from Ref.~\cite{Sang:2023ncm} without this factor.

For $T_{4b}$, we use the matrix elements in Table~III of
Ref.~\cite{Yang:2026bjy}. In the scalar channel, Sets~3 and~4
correspond to the two orthogonal states $0_{\mathrm{I}}^{++}$ and
$0_{\mathrm{II}}^{++}$ defined in Eq.~(14) of that reference, with
$\theta_b=66^\circ$. We absorb the mixing factors into the scalar
LDMEs. For Set~3, the sextet, antitriplet, and mixed matrix elements
are multiplied by $\sin^2\theta_b$, $\cos^2\theta_b$, and
$\sin\theta_b\cos\theta_b$, respectively. For Set~4, the two
diagonal factors are interchanged and the mixed matrix element
changes sign. In Eq.~(15) of Yang \textit{et al.}~\cite{Yang:2026bjy}, however,
the diagonal weights are the same for both scalar $T_{4c}$ states.
For $J=1,2$, both sets use the $T_{4b}$ LDMEs from Table~III
directly, since they already include $2J+1$.

For Set~5, we take the radial wave functions at the origin from
Eqs.~(27) and~(28) of Ref.~\cite{Liang:2025wbt} and use the same
values for all three spin channels. In the diquark--antidiquark
model~\cite{Feng:2020qee,Huang:2021vtb}, the
$\bar{\mathbf{3}}\otimes\mathbf{3}$ LDME is then estimated as
\begin{equation}
	\langle\Othree^{(J)}\rangle_{T_{4b}}
	\approx \frac{2J+1}{4\pi^3}
	|R_{bb}(0)|^4 |R_{d\bar d}(0)|^2,
	\label{eq:bottom-radial-LDME}
\end{equation}
where $R_{bb}$ and $R_{d\bar d}$ are the radial wave functions
for the internal motion of the diquark $d=(bb)_{\bar{\mathbf{3}}}$
and its motion relative to the antidiquark
$\bar d=(\bar b\bar b)_{\mathbf{3}}$, respectively.
Using $|R_{bb}(0)|^2=1.810~\mathrm{GeV}^3$ and
$|R_{d\bar d}(0)|^2=11.739~\mathrm{GeV}^3$, we obtain
$0.3101$, $0.9303$, and $1.5504~\mathrm{GeV}^9$ for $J=0,1,2$,
respectively. Since Liang \textit{et al.} do not provide the sextet
or mixed LDME, Set~5 determines only the
$\bar{\mathbf{3}}\otimes\mathbf{3}$ contribution to the scalar
decay width.

The radiative branching fractions are given by
\begin{equation}
	\mathcal{B}(\TQ\to H+\gamma)
	= \frac{\Gamma(\TQ\to H+\gamma)}
	{\Gamma_{\mathrm{tot}}(\TQ)}.
	\label{eq:branching-fraction}
\end{equation}
To estimate the branching fractions, we use a common reference
width for all states within each flavor sector. For $T_{4c}$,
we adopt $120~\mathrm{MeV}$ following
Refs.~\cite{Sang:2023ncm,ATLAS:2022hhx}. For $T_{4b}$, we use
the $9.62~\mathrm{MeV}$ from Ref.~\cite{Agaev:2023ara}.
\begin{table*}[t]
	\centering
	\caption{LO and NLO predictions for $T_{4c}$ radiative decays.
		Decay widths are in MeV, and branching fractions use
		$\Gamma_{\mathrm{tot}}(T_{4c})=120~\mathrm{MeV}$.
		Central values are evaluated at $\mu_R=2m_c$.
		The uncertainties are estimated by varying the renormalization
		scale $\mu_R$ from $m_c$ to $4m_c$.
	}
	\label{tab:T4c-widths}
	\begingroup
	\small
	\renewcommand{\arraystretch}{1.20}
	\setlength{\tabcolsep}{4pt}
	\resizebox{\textwidth}{!}{%
		\begin{tabular}{|c|c|*{10}{c|}}
			\hline
			\multicolumn{12}{|l|}{$m_c=1.5\,\mathrm{GeV}
				\qquad \mu_R=2m_c
				\qquad \Gamma\;(\mathrm{MeV})$} \\
			\hline
			\multirow{2}{*}{Channel}
			& \multirow{2}{*}{LDME set}
			& \multicolumn{5}{c|}{LO}
			& \multicolumn{5}{c|}{NLO} \\
			\cline{3-12}
			& & $\mathbf6\otimes\bar{\mathbf6}$
			& $\bar{\mathbf3}\otimes\mathbf3$
			& mix & Total & $\mathcal Br$
			& $\mathbf6\otimes\bar{\mathbf6}$
			& $\bar{\mathbf3}\otimes\mathbf3$
			& mix & Total & $\mathcal Br$ \\
			\hline
			
			\multirow{2}{*}{$T_{4c}^{0^{++}}\to J/\psi+\gamma$}
			& Set~1
			& $0.003^{+0.005}_{-0.002}$
			& $0.596^{+0.999}_{-0.318}$
			& $-0.086^{+0.046}_{-0.144}$
			& $0.513^{+0.860}_{-0.274}$
			& $4.276^{+7.165}_{-2.284}\times10^{-3}$
			& $-0.003^{+0.003}_{-0.018}$
			& $1.545^{+1.530}_{-0.703}$
			& $-0.076^{+0.135}_{+0.011}$
			& $1.466^{+1.647}_{-0.689}$
			& $1.222^{+1.372}_{-0.574}\times10^{-2}$ \\
			\cline{2-12}
			& Set~2
			& $0.003^{+0.006}_{-0.002}$
			& $0.321^{+0.538}_{-0.172}$
			& $0.066^{+0.110}_{-0.035}$
			& $0.390^{+0.654}_{-0.208}$
			& $3.252^{+5.449}_{-1.737}\times10^{-3}$
			& $-0.003^{+0.003}_{-0.019}$
			& $0.832^{+0.824}_{-0.379}$
			& $0.058^{-0.009}_{-0.103}$
			& $0.888^{+0.702}_{-0.384}$
			& $7.399^{+5.853}_{-3.203}\times10^{-3}$ \\
			\hline
			
			\multirow{2}{*}{$T_{4c}^{2^{++}}\to J/\psi+\gamma$}
			& Set~1
			& \textemdash
			& $0.023^{+0.039}_{-0.012}$
			& \textemdash
			& $0.023^{+0.039}_{-0.012}$
			& $1.935^{+3.242}_{-1.033}\times10^{-4}$
			& \textemdash
			& $0.068^{+0.079}_{-0.032}$
			& \textemdash
			& $0.068^{+0.079}_{-0.032}$
			& $5.684^{+6.589}_{-2.695}\times10^{-4}$ \\
			\cline{2-12}
			& Set~2
			& \textemdash
			& $0.020^{+0.034}_{-0.011}$
			& \textemdash
			& $0.020^{+0.034}_{-0.011}$
			& $1.693^{+2.837}_{-0.904}\times10^{-4}$
			& \textemdash
			& $0.060^{+0.069}_{-0.028}$
			& \textemdash
			& $0.060^{+0.069}_{-0.028}$
			& $4.974^{+5.765}_{-2.358}\times10^{-4}$ \\
			\hline
			
			\multirow{2}{*}{$T_{4c}^{1^{+-}}\to\eta_c+\gamma$}
			& Set~1
			& \textemdash
			& $0.180^{+0.301}_{-0.096}$
			& \textemdash
			& $0.180^{+0.301}_{-0.096}$
			& $1.496^{+2.506}_{-0.799}\times10^{-3}$
			& \textemdash
			& $0.399^{+0.301}_{-0.171}$
			& \textemdash
			& $0.399^{+0.301}_{-0.171}$
			& $3.327^{+2.507}_{-1.426}\times10^{-3}$ \\
			\cline{2-12}
			& Set~2
			& \textemdash
			& $0.110^{+0.185}_{-0.059}$
			& \textemdash
			& $0.110^{+0.185}_{-0.059}$
			& $9.206^{+15.424}_{-4.917}\times10^{-4}$
			& \textemdash
			& $0.246^{+0.185}_{-0.105}$
			& \textemdash
			& $0.246^{+0.185}_{-0.105}$
			& $2.047^{+1.543}_{-0.878}\times10^{-3}$ \\
			\hline
		\end{tabular}%
	}
	\endgroup
\end{table*}
\begin{table*}[t]
	\centering
	\caption{LO and NLO predictions for $T_{4b}$ radiative decays.
		Decay widths are in eV, and branching fractions use
		$\Gamma_{\mathrm{tot}}(T_{4b})=9.62~\mathrm{MeV}$.
		Central values are evaluated at $\mu_R=2m_b$.
		The uncertainties are estimated by varying the renormalization
		scale $\mu_R$ from $m_b$ to $4m_b$.}
	\label{tab:T4b-widths}
	\begingroup
	\small
	\renewcommand{\arraystretch}{1.20}
	\setlength{\tabcolsep}{4pt}
	\resizebox{\textwidth}{!}{%
		\begin{tabular}{|c|c|*{10}{c|}}
			\hline
			\multicolumn{12}{|l|}{$m_b=4.9\,\mathrm{GeV}
				\qquad \mu_R=2m_b
				\qquad \Gamma\;(\mathrm{eV})$} \\
			\hline
			\multirow{2}{*}{Channel}
			& \multirow{2}{*}{LDME set}
			& \multicolumn{5}{c|}{LO}
			& \multicolumn{5}{c|}{NLO} \\
			\cline{3-12}
			& & $\mathbf6\otimes\bar{\mathbf6}$
			& $\bar{\mathbf3}\otimes\mathbf3$
			& mix & Total & $\mathcal Br$
			& $\mathbf6\otimes\bar{\mathbf6}$
			& $\bar{\mathbf3}\otimes\mathbf3$
			& mix & Total & $\mathcal Br$ \\
			\hline
			
			\multirow{2}{*}{$T_{4b}^{0^{++}}\to\Upsilon+\gamma$}
			& Set~3
			& $0.395^{+0.353}_{-0.167}$
			& $1.203^{+1.073}_{-0.509}$
			& $-1.379^{+0.583}_{-1.230}$
			& $0.220^{+0.196}_{-0.093}$
			& $2.284^{+2.038}_{-0.966}\times10^{-8}$
			& $-0.196^{+0.244}_{-0.886}$
			& $2.465^{+1.039}_{-0.780}$
			& $-1.071^{+0.949}_{+0.021}$
			& $1.198^{+1.102}_{-0.514}$
			& $1.246^{+1.145}_{-0.535}\times10^{-7}$ \\
			\cline{2-12}
			& Set~4
			& $0.078^{+0.070}_{-0.033}$
			& $6.065^{+5.411}_{-2.565}$
			& $1.379^{+1.230}_{-0.583}$
			& $7.522^{+6.712}_{-3.182}$
			& $7.819^{+6.977}_{-3.308}\times10^{-7}$
			& $-0.039^{+0.048}_{-0.176}$
			& $12.429^{+5.236}_{-3.931}$
			& $1.071^{-0.021}_{-0.949}$
			& $13.461^{+4.112}_{-3.904}$
			& $1.399^{+0.427}_{-0.406}\times10^{-6}$ \\
			\hline
			
			\multirow{2}{*}{$T_{4b}^{2^{++}}\to\Upsilon+\gamma$}
			& Set~3
			& \textemdash
			& $0.684^{+0.610}_{-0.289}$
			& \textemdash
			& $0.684^{+0.610}_{-0.289}$
			& $7.108^{+6.343}_{-3.007}\times10^{-8}$
			& \textemdash
			& $1.594^{+0.825}_{-0.539}$
			& \textemdash
			& $1.594^{+0.825}_{-0.539}$
			& $1.657^{+0.858}_{-0.560}\times10^{-7}$ \\
			\cline{2-12}
			& Set~5
			& \textemdash
			& $0.663^{+0.592}_{-0.280}$
			& \textemdash
			& $0.663^{+0.592}_{-0.280}$
			& $6.892^{+6.150}_{-2.915}\times10^{-8}$
			& \textemdash
			& $1.546^{+0.800}_{-0.523}$
			& \textemdash
			& $1.546^{+0.800}_{-0.523}$
			& $1.607^{+0.832}_{-0.543}\times10^{-7}$ \\
			\hline
			
			\multirow{2}{*}{$T_{4b}^{1^{+-}}\to\eta_b+\gamma$}
			& Set~3
			& \textemdash
			& $2.927^{+2.612}_{-1.238}$
			& \textemdash
			& $2.927^{+2.612}_{-1.238}$
			& $3.043^{+2.715}_{-1.287}\times10^{-7}$
			& \textemdash
			& $5.359^{+1.748}_{-1.580}$
			& \textemdash
			& $5.359^{+1.748}_{-1.580}$
			& $5.571^{+1.817}_{-1.642}\times10^{-7}$ \\
			\cline{2-12}
			& Set~5
			& \textemdash
			& $2.839^{+2.534}_{-1.201}$
			& \textemdash
			& $2.839^{+2.534}_{-1.201}$
			& $2.952^{+2.634}_{-1.249}\times10^{-7}$
			& \textemdash
			& $5.199^{+1.695}_{-1.532}$
			& \textemdash
			& $5.199^{+1.695}_{-1.532}$
			& $5.404^{+1.762}_{-1.593}\times10^{-7}$ \\
			\hline
		\end{tabular}%
	}
	\endgroup
\end{table*}

Tables~\ref{tab:T4c-widths} and~\ref{tab:T4b-widths} summarize the
radiative decay widths and branching fractions. For
$T_{4Q}^{0^{++}}\to V_Q+\gamma$, the contributions from the two
color configurations and their interference are shown separately.
At the central scale $\mu_R=2m_Q$, the NLO QCD corrections enhance
the decay widths in all channels. The scale uncertainties are
estimated by varying $\mu_R$ from $m_Q$ to $4m_Q$. We define
$K=\Gamma_{\mathrm{NLO}}/\Gamma_{\mathrm{LO}}$ using the same inputs
at both orders. For $T_{4Q}^{2^{++}}\to V_Q+\gamma$ and
$T_{4Q}^{1^{+-}}\to\eta_Q+\gamma$, the $K$ factors are $2.94$ and
$2.22$ for $T_{4c}$, and $2.33$ and $1.83$ for $T_{4b}$, respectively.
Only one tetraquark LDME contributes to each of these channels, so
it cancels in the ratio.

For $T_{4c}^{0^{++}}\to J/\psi+\gamma$, the dominant
$\bar{\mathbf{3}}\otimes\mathbf{3}$ contribution increases by a
factor of $2.59$ at NLO in both sets, while the magnitude of the
interference term decreases by about $12\%$. At both LO and NLO,
the interference is destructive for Set~1 and constructive for Set~2.
The resulting NLO widths are $1.466~\mathrm{MeV}$ ($K=2.86$) for
Set~1 and $0.888~\mathrm{MeV}$ ($K=2.27$) for Set~2. For
$T_{4b}^{0^{++}}\to\Upsilon+\gamma$, destructive interference in
Set~3 suppresses the LO width to $0.220~\mathrm{eV}$. The NLO width
is $1.198~\mathrm{eV}$, giving $K=5.45$. In Set~4, the interference
is constructive, and the width increases from $7.522~\mathrm{eV}$
at LO to $13.461~\mathrm{eV}$ at NLO ($K=1.79$).
For both flavors, the diagonal $\mathbf{6}\otimes\bar{\mathbf{6}}$
contribution to $T_{4Q}^{0^{++}}\to V_Q+\gamma$ becomes negative
at NLO because the negative one-loop correction is larger in magnitude
than the LO term. The physical partial widths remain positive after
summing the two diagonal contributions and their interference.

At LO, the decay widths depend on the renormalization scale only
through $\alpha_s^4(\mu_R)$. At $\mu_R=m_Q$, the $T_{4c}$ decay
widths are about $168\%$ above their central values, and the
$T_{4b}$ decay widths are about $89\%$ above theirs. At
$\mu_R=4m_Q$, they are about $53\%$ and $42\%$ below their
central values, respectively. The relative scale dependence is
reduced at NLO in most channels. For
$T_{4c}^{2^{++}}\to J/\psi+\gamma$, the NLO decay width is
$116\%$ above its central value at $\mu_R=m_c$ and $47\%$ below
it at $\mu_R=4m_c$. For $T_{4b}^{2^{++}}\to\Upsilon+\gamma$,
the corresponding increase and decrease are $52\%$ and $34\%$.
An exception is $T_{4b}^{0^{++}}\to\Upsilon+\gamma$ with Set~3,
where the NLO decay width is $92\%$ above its central value at
$\mu_R=m_b$ and $43\%$ below it at $\mu_R=4m_b$. Both variations
are slightly larger than at LO. The sizable NLO corrections and
the residual scale dependence suggest that higher-order QCD
corrections may still be important.

We estimate the number of $T_{4c}\to H+\gamma$ decays in $pp$ collisions at
$\sqrt{s}=13~\mathrm{TeV}$ for an integrated luminosity of
$\mathcal L=100~\mathrm{fb}^{-1}$ using
\begin{equation}
	N_{\mathrm{events}}
	=\mathcal L\,\sigma_{\mathrm{LHC}}\,
	\mathcal Br_{\mathrm{NLO}}(T_{4c}\to H+\gamma).
	\label{eq:radiative-event-yield}
\end{equation}
We take the LO production cross sections from Table~II of
Ref.~\cite{Feng:2023agq}, integrated over
$6~\mathrm{GeV}\leq p_T\leq100~\mathrm{GeV}$ and $|y|\leq5$
for the produced $T_{4c}$. We combine the Model~I and Model~II
cross sections with the NLO branching fractions for Sets~1
and~2, respectively, from Table~\ref{tab:T4c-widths}.
The resulting event yields are listed in
Table~\ref{tab:T4c-event-yields}. The first uncertainty comes
from the quoted symmetric scale uncertainty on the production
cross section and the second from varying the renormalization
scale $\mu_R$ from $m_c$ to $4m_c$ in the decay calculation.
In $T_{4c}^{0^{++}}$ production, the two color configurations interfere
constructively in Model~I and destructively in Model~II.
The corresponding central cross sections are $37~\mathrm{nb}$
and $9~\mathrm{nb}$, respectively~\cite{Feng:2023agq}.
With NLO branching fractions of $1.22\%$ and $0.740\%$,
the central yield for Set~1 is $4.52\times10^7$ decays,
about $6.8$ times the Set~2 yield of $6.66\times10^6$.
\begin{table*}[t]
	\centering
	\caption{Expected numbers of $T_{4c}\to H+\gamma$ decays
		($H=J/\psi,\eta_c$) in $pp$ collisions at
		$\sqrt{s}=13~\mathrm{TeV}$ for $\mathcal L=100~\mathrm{fb}^{-1}$.
		We combine the LO production cross sections from
		Ref.~\cite{Feng:2023agq}, integrated over
		$6~\mathrm{GeV}\leq p_T\leq100~\mathrm{GeV}$ and $|y|\leq5$,
		with the NLO decay branching fractions.
		The first uncertainty comes from the production cross section
		and the second from varying the renormalization scale $\mu_R$
		from $m_c$ to $4m_c$ in the decay calculation.}
	\label{tab:T4c-event-yields}
	\begingroup
	\footnotesize
	\renewcommand{\arraystretch}{1.35}
	\setlength{\tabcolsep}{12pt}
	\resizebox{\textwidth}{!}{%
		\begin{tabular}{|c|c|c|c|c|}
			\hline
			\multirow{2}{*}{Channel}
			& \multicolumn{2}{c|}{Set~1}
			& \multicolumn{2}{c|}{Set~2} \\
			\cline{2-5}
			& $\sigma_{\mathrm{LHC}}~(\mathrm{nb})$
			& $N_{\mathrm{events}}$
			& $\sigma_{\mathrm{LHC}}~(\mathrm{nb})$
			& $N_{\mathrm{events}}$ \\
			\hline
			
			$T_{4c}^{0^{++}}\to J/\psi+\gamma$
			& $37\pm26$
			& $\left(4.520\pm3.177{}^{+5.078}_{-2.125}\right)
			\times10^{7}$
			& $9\pm6$
			& $\left(6.659\pm4.439{}^{+5.268}_{-2.883}\right)
			\times10^{6}$ \\
			\hline
			
			$T_{4c}^{2^{++}}\to J/\psi+\gamma$
			& $93\pm65$
			& $\left(5.286\pm3.695{}^{+6.128}_{-2.506}\right)
			\times10^{6}$
			& $81\pm57$
			& $\left(4.029\pm2.835{}^{+4.670}_{-1.910}\right)
			\times10^{6}$ \\
			\hline
			
			$T_{4c}^{1^{+-}}\to\eta_c+\gamma$
			& $0.28\pm0.16$
			& $\left(9.315\pm5.323{}^{+7.020}_{-3.994}\right)
			\times10^{4}$
			& $0.17\pm0.10$
			& $\left(3.480\pm2.047{}^{+2.623}_{-1.492}\right)
			\times10^{4}$ \\
			\hline
		\end{tabular}%
	}
	\endgroup
\end{table*}
For both LDME sets, the central yields are of order
$10^6$ for $T_{4c}^{2^{++}}\to J/\psi+\gamma$ and
$10^4$ for $T_{4c}^{1^{+-}}\to\eta_c+\gamma$.
Although $T_{4c}^{1^{+-}}$ has a larger radiative branching
fraction, its much smaller production cross section leads
to fewer events.
The $T_{4c}\to J/\psi+\gamma$ channels can be searched for
in the $\ell^+\ell^-\gamma$ invariant-mass spectrum,
with the $J/\psi$ reconstructed through
$J/\psi\to\ell^+\ell^-$ ($\ell=e,\mu$).
Such searches would complement those in di-charmonium
final states.
\section{Summary}\label{sec:summary}
We have calculated the NLO QCD corrections to
$T_{4Q}^{0^{++}}\to V_Q+\gamma$, $T_{4Q}^{2^{++}}\to V_Q+\gamma$,
and $T_{4Q}^{1^{+-}}\to\eta_Q+\gamma$. We obtain analytic expressions
for all five SDCs, including the interference coefficient for
$T_{4Q}^{0^{++}}\to V_Q+\gamma$. After renormalization and matching,
the divergences cancel in all five SDCs, as required by NRQCD
factorization at this order. For $T_{4Q}^{2^{++}}\to V_Q+\gamma$
and $T_{4Q}^{1^{+-}}\to\eta_Q+\gamma$, the LDMEs cancel in the
ratio $K=\Gamma_{\mathrm{NLO}}/\Gamma_{\mathrm{LO}}$. At the
central scale, the $K$ factors for these two channels are $2.94$
and $2.22$ for $T_{4c}$, and $2.33$ and $1.83$ for $T_{4b}$,
respectively. For $T_{4c}^{0^{++}}\to J/\psi+\gamma$, the
$\bar{\mathbf{3}}\otimes\mathbf{3}$ coefficient increases by a
factor of $2.59$ at NLO, while the magnitude of the interference
coefficient decreases by about $12\%$.

The NLO corrections reduce $\mu_R$ dependence
of the decay widths in most channels. For $T_{4b}^{0^{++}}\to\Upsilon+\gamma$ with Set~3,
however, the scale dependence is slightly stronger at NLO
than at LO. Taking $\Gamma_{\mathrm{tot}}(T_{4c})=120~\mathrm{MeV}$,
we combine the NLO branching fractions with the LO production
cross sections to estimate the decay yields in $pp$ collisions
at $\sqrt{s}=13~\mathrm{TeV}$. For an integrated luminosity of
$100~\mathrm{fb}^{-1}$ and the cuts in Table~\ref{tab:T4c-event-yields},
the expected yields are of order $10^6$--$10^7$ for
$T_{4c}^{0^{++}}\to J/\psi+\gamma$, $10^6$ for
$T_{4c}^{2^{++}}\to J/\psi+\gamma$, and $10^4$ for
$T_{4c}^{1^{+-}}\to\eta_c+\gamma$. The $J/\psi+\gamma$ channels
can be studied in the $\ell^+\ell^-\gamma$ invariant-mass spectrum
through $J/\psi\to\ell^+\ell^-$, complementing searches in
di-charmonium final states.

\medskip
\noindent\textit{Note added.}
While this manuscript was being finalized, a study of fully
charmed tetraquark radiative decays at NLO in QCD
appeared~\cite{Li:2026wvs}, in which the authors reached
conclusions broadly similar to ours.
Our calculation was performed independently and also includes
the corresponding fully bottom tetraquark decays.

\section*{Acknowledgments}
The work of Y.-D. Z. and Z.-F. J. is supported by the Young and Middle-aged Scientific and Technological Innovation Team Program for Colleges and Universities of the Hubei Provincial Department of Education under Grant No. T2024029 and by the 2025 Provincial Teaching Reform Research Project of Hubei Provincial
Department of Education under Grant No. 2025468.

\appendix
\section{Analytic NLO coefficients}
\label{app:finite-NLO-coefficients}
Here we collect the analytic expressions for the finite NLO
coefficients $c_i^{(1)}$ defined in Eq.~\eqref{eq:NLO-scale},
\begin{widetext}
	\raggedbottom
	\allowdisplaybreaks[4]

	\begin{subequations}\label{eq:finite-NLO-coefficients}
		\begin{align}
			c_1^{(1)}
			&=\frac{c_1^{(0)}}{3231900}\Biggl\{
			\frac{3556479443}{126}\mathcal L_4
			+\frac{4744600}{9}\mathcal L_7
			+\frac{162976643080}{9009}\mathcal L_{10}
			+535080\mathcal L_{28}
			\nonumber\\
			&\quad
			+\frac{111356875}{78}\mathcal Q
			+10\biggl(\frac{9150200363}{3003}+34020n_l\biggr)\mathcal F
			+2\biggl(311283n_h
			\nonumber\\
			&\qquad
			-\frac{20022539963}{1001}\biggr)\mathcal H
			+210(27lbl-5)\mathcal G
			+\ell_2\biggl(
			\frac{1180128114988}{9009}\ell_2
			-\frac{26323401155}{3003}\ell_5
			\nonumber\\
			&\qquad
			+\frac{1556267647}{7}
			-21230748n_h-4572288n_l
			\biggr)
			+15\ell_3\biggl(5355\ell_2+\frac{30567898}{7}\biggr)
			\nonumber\\
			&\quad
			+10\ell_5\biggl(
			531765-\frac{27020132}{231}\ell_5-37422n_l
			\biggr)
			-22965705\mathcal U
			-\biggl(11265167
			\nonumber\\
			&\qquad
			+\frac{591948n_h}{5}\biggr)\mathcal T
			+\frac{90315918892}{9009}\pi^2
			+216(136237+58639n_h-17192n_l)
			\Biggr\},
			\label{eq:f1}\\[0.8ex]
			c_2^{(1)}
			&=\frac{c_2^{(0)}}{33346404}\Biggl\{
			\frac{958215805}{126}\mathcal L_4
			-\frac{61431832}{9}\mathcal L_7
			+\frac{151860204812}{9009}\mathcal L_{10}
			+123480\mathcal L_{28}
			\nonumber\\
			&\quad
			-\frac{4454275}{78}\mathcal Q
			+2\biggl(\frac{8114939317}{1001}-170100n_l\biggr)\mathcal F
			-2\biggl(311283n_h
			\nonumber\\
			&\qquad
			+\frac{3324629617}{1001}\biggr)\mathcal H
			-42(135lbl+1937)\mathcal G
			+\ell_2\biggl(
			\frac{463889734052}{9009}\ell_2
			\nonumber\\
			&\qquad
			-\frac{19369184657}{1001}\ell_5
			+\frac{8593736429}{7}
			-210758436n_h-60963840n_l
			\biggr)
			\nonumber\\
			&\quad
			+3\ell_3\biggl(\frac{636365918}{35}-1071\ell_2\biggr)
			+2\ell_5\biggl(
			\frac{243218678}{77}\ell_5-7693785+295974n_l
			\biggr)
			\nonumber\\
			&\quad
			-54408375\mathcal U
			+\biggl(1956707+\frac{4919292n_h}{25}\biggr)\mathcal T
			+\frac{46137336478}{1001}\pi^2
			\nonumber\\
			&\quad
			+216\biggl(
			1309919+\frac{2704261n_h}{5}-170968n_l
			\biggr)
			\Biggr\},
			\label{eq:f2}\\[0.8ex]
			c_{\mathrm{mix}}^{(1)}
			&=\frac{c_{\mathrm{mix}}^{(0)}}{2}
			\left(
			\frac{c_1^{(1)}}{c_1^{(0)}}
			+\frac{c_2^{(1)}}{c_2^{(0)}}
			\right),
			\label{eq:fmix}\\[0.8ex]
			c_3^{(1)}
			&=\frac{c_3^{(0)}}{38346479892}\Biggl\{
			-\frac{7232031048397}{126}\mathcal L_4
			-\frac{193339704908}{9}\mathcal L_7
			+\frac{37026781209311}{819}\mathcal L_{10}
			\nonumber\\
			&\quad
			-\frac{179166870575}{78}\mathcal Q
			+3\biggl(
			\frac{234320239099}{91}-1455375600n_l
			\biggr)\mathcal F
			-2\biggl(
			\frac{11874314209315}{91}
			\nonumber\\
			&\qquad
			+3391525242n_h+15580350lbl
			\biggr)\mathcal H
			-3(58483007+9150030lbl)\mathcal G
			\nonumber\\
			&\quad
			+135(1458000lbl-1044061967)\mathcal U
			+\ell_2\biggl(
			\frac{466269004662224}{819}\ell_2
			-\frac{14837361715937}{182}\ell_5
			\nonumber\\
			&\qquad
			+1575031079191
			-163337506668n_h-67474669248n_l
			+4007920500lbl
			\biggr)
			\nonumber\\
			&\quad
			-3\ell_3\biggl(
			6912641043\ell_2+\frac{156101342288}{5}
			\biggr)
			+2\ell_5\biggl(
			\frac{107483454884}{7}\ell_5
			\nonumber\\
			&\qquad
			-24491806065+3135558762n_l
			\biggr)
			+\biggl(
			\frac{39131089956n_h}{25}
			+290681460lbl
			\nonumber\\
			&\qquad
			-29814980873
			\biggr)\mathcal T
			+\frac{472615186329865}{1638}\pi^2
			+36\biggl(
			8357573955+\frac{8231349938n_h}{5}
			\nonumber\\
			&\qquad
			-1157588432n_l-27473000lbl
			\biggr)
			\Biggr\},
			\label{eq:f3}\\[0.8ex]
			c_4^{(1)}
			&=\frac{c_4^{(0)}}{14948388}\Biggl\{
			\frac{2260066553}{1512}\mathcal L_4
			-\frac{54321134}{27}\mathcal L_7
			+\frac{189155723083}{54054}\mathcal L_{10}
			+22890\mathcal L_{28}
			\nonumber\\
			&\quad
			-\frac{75279575}{936}\mathcal Q
			+16\biggl(
			\frac{1547586419}{9009}-10395n_l
			\biggr)\mathcal F
			-\biggl(
			\frac{151376214191}{12012}
			\nonumber\\
			&\qquad
			+267624n_h+317025lbl
			\biggr)\mathcal H
			-\frac58(24791+2304lbl)\mathcal G
			+\ell_2\biggl(
			\frac{713840544652}{27027}\ell_2
			\nonumber\\
			&\qquad
			-\frac{57821799142}{9009}\ell_5
			+\frac{68998499029}{168}
			-42183666n_h-22212036n_l
			\nonumber\\
			&\qquad
			+163350lbl
			\biggr)
			+\ell_3\biggl(
			\frac{237545047}{70}-\frac{3570183}{4}\ell_2
			\biggr)
			+\ell_5\biggl(
			\frac{370528623}{154}\ell_5
			\nonumber\\
			&\qquad
			-4623490+264222n_l
			\biggr)
			+\frac34(129600lbl-14286257)\mathcal U
			+\biggl(
			\frac{1810998n_h}{25}
			\nonumber\\
			&\qquad
			+11070lbl-\frac{4852781}{24}
			\biggr)\mathcal T
			+\frac{2741882598487}{162162}\pi^2
			+\frac{178865151633}{4394}
			\nonumber\\
			&\quad
			+28\biggl(
			\frac{2646538n_h}{5}-196285n_l
			\biggr)
			+29700lbl
			\Biggr\}.
			\label{eq:f4}
		\end{align}
	\end{subequations}

	The common real quantities are defined by
	\nopagebreak[4]
	\begin{subequations}\label{eq:NLO-abbreviations}
		\begin{align}
			\mathcal L_n
			&=\operatorname{Re}\operatorname{Li}_2(n),
			\qquad
			\ell_2=\ln2,
			\qquad
			\ell_3=\ln3,
			\qquad
			\ell_5=\ln5,
			\label{eq:NLO-dilog}\\
			\omega
			&=\ln(2+\sqrt3),
			\qquad
			\vartheta=\arccos\!\left(-\frac14\right),
			\qquad
			\mathcal H=\ell_2^2+\vartheta^2,
			\qquad
			\mathcal T=\sqrt{15}\,\vartheta,
			\label{eq:NLO-logs}\\
			\mathcal U
			&=\omega^2+\pi^2,
			\qquad
			\mathcal F=\mathcal L_5+\frac12\ell_2\ell_5-\frac{5\pi^2}{12},
			\qquad
			\mathcal Q=\mathcal L_{96}+\ln95\ln96-\frac{\pi^2}{6},
			\label{eq:NLO-HF}\\
			\mathcal G
			&=120\mathcal U-230\mathcal H+2535\ell_2
			-840\sqrt3\,\omega-33\mathcal T+60.
			\label{eq:NLO-G}
		\end{align}
	\end{subequations}

\end{widetext}
Here $n_l$ counts massless quark flavors, $n_h$ counts heavy
quark flavors of mass $m_Q$, and $lbl$ labels the
light-by-light contributions.

\end{document}